\documentclass[a4paper,11pt]{article}
\usepackage{amsmath}
\usepackage{amssymb}
\usepackage{graphicx}
\usepackage{caption}
\usepackage[sort]{cite}
\usepackage{subcaption}
\usepackage{cite}
\usepackage[T1]{fontenc}
\usepackage[a4paper,left=1.0in, right=1.0in,top=1.1in, bottom=1.2in]{geometry}
\usepackage[affil-it]{authblk}
\newcommand{\be}{\begin{equation}}
\newcommand{\ee}{\end{equation}}

\begin{document}
\titlepage
\vspace*{1in}
\begin{center}
{\Large \bf The rapidity dependence of the average transverse momentum \\ in p+p and p+Pb collisions - revisited}\\
\vspace*{0.5cm}
Katarzyna Deja\footnote{e-address: katarzyna.deja@ncbj.gov.pl} \\
\vspace*{0.5cm}
 {\it Narodowe Centrum Bada\'n J\k{a}drowych, Andrzeja So\l tana 7, 05-400 Otwock, Polska\\}
\vspace*{0.5cm}
Krzysztof Kutak \footnote{e-address: krzysztof.kutak@ifj.edu.pl}\\
\vspace*{0.5cm}
 {\it Instytut Fizyki J\k{a}drowej Polskiej Akademii Nauk, Radzikowskiego 152, 31-342 Krak\' ow, Polska\\}
\end{center}
\vspace*{1cm}
\centerline{(\today)}

\vskip1cm
\begin{abstract}
We revisit the calculation of rapidity dependence of the average transverse momentum $\left\langle p_{T}\right\rangle$ of single inclusive jet production or particle in a $p$+$p$ and $p$+$Pb$  collision, using the high energy factorization. We update previous predictions for the $\left\langle p_{T}\right\rangle$ both in central and forward rapidity region using parton densities following from extended Balitsky-Kovchegov and Balitsky-Fadin-Kuraev-Lipatov evolution equations to account for corrections of higher orders as well as one obtained by Sudakov resummation. Furthermore, we demonstrate that in the midrapidity region the saturation based formalisms predict increase of the transversal momentum while in the forward rapidity region the $\left\langle p_{T}(y)\right\rangle$ decreases.
\end{abstract}

\section*{Introduction}

Over the past decades, the understanding of QCD at the regime of high energy and high density is gradually advancing. In particular the basic equations of formalism, called color glass condensate (CGC), for review see \cite{Gelis:2010nm}, allowing for calculations of observables when the system of partons is dense is well established (it has been recently advanced to account for corrections of higher orders \cite{Lublinsky:2016meo,Balitsky:2008zza,Lappi:2015fma}). For certain observables (and when some simplifications are assumed) the color glass condensate formalism can be easily linked to the momentum space formulated high energy factorization (HEF) \cite{Catani:1990eg,Collins:1991ty}. The cross section for observables in the framework of high energy factorization is calculated as a convolution of appropriate hard matrix elements, with parton densities which parametrize colliding hadrons. In general, depending on kinematical setup, the partons in the initial state are probed at large, moderate or low longitudinal momentum fraction of parent hadron.\\
In the case when the final state is at forward rapidity the formula for the cross section is provided by the so-called hybrid \cite{Dumitru:2005gt,Deak:2009xt,Deak:2009ae,Deak:2010gk} version of HEF. In such an approach partons in one of the incoming hadrons carry large a longitudinal momentum fraction while in the other one a small longitudinal momentum fraction. If the longitudinal momentum fraction of partons $x$ is low enough the QCD framework predicts effects of gluon saturation \cite{Gribov:1984tu} and indeed there are hints that saturation occurs \cite{GolecBiernat:1998js,Albacete:2010pg,Dumitru:2010iy,Dusling:2013qoz,Kutak:2012rf,Chatrchyan:2008aa}.
An interesting idea is to test for some simple observable predictions based on the assumption that saturation exist with another completely different framework. In particular in \cite{Bozek:2013sda} the authors compare results for single inclusive hadron production (see also \cite{Levin:2010dw}) as obtained in the simple model having bulk properties of CGC versus hydrodynamics (hydrodynamical model of heavy ion collisions). In particular they focus on average transversal momentum for single inclusive hadron production. In the cited paper the authors used HEF in the midrapidity regime with a model for gluon density and concluded that saturation effects manifest themselves by increasing mean $p_T$ of hadrons or jets as one goes away from the certain rapidity region towards the forward rapidity region (following the direction of proton). The calculation in principle was justified around the central rapidity region where the formalism is applicable. Furthermore their results obtained from hydrodynamical calculation have shown an opposite trend, i.e. decreasing cross section as one goes to the forward rapidity region. So they concluded that a study of the mean $p_T$ of hadrons can provide a handle on what is realized in nature hydrodynamics or CGC. However, the study of the observable was repeated by other authors using other parton density sets and formalism applicable at the forward rapidity region \cite{Duraes:2015qoa}. Furthermore the result of \cite{Duraes:2015qoa} shows strong dependence on whether one addresses individual particles or jets. The authors of \cite{Duraes:2015qoa} arrived at the conclusion that for particles the cross section decreases in the whole rapidity while for jets it increases in the moderate rapidity region and then decreases in the large rapidity region. In this paper in order to shed light on the apparent discrepancy we revisit both QCD calculations in their range of applicability using gluon densities \cite{Kutak:2012rf} obtained from extending Balitskii-Fadin-Kuraev-Lipatov (BFKL) \cite{Kuraev:1977fs,Balitsky:1978ic,Kuraev:1976ge}, and Balitsky-Kovchegov (BK) \cite{Balitsky:1995ub,Kovchegov:1999yj} evolution equations as well as from Sudakov based resummation [26]. Furthermore we aim at understanding better the role of saturation for specific observable, i.e. jet vs. hadron.

\section{Rapidity dependence of average transverse momentum}
\label{sec-form-general}

The single inclusive jet production process can be schematically written as
$$
\label{proccess}
A+B \longrightarrow a+b \longrightarrow jet + X, 
$$
where A and B are colliding hadrons, each of which provides a parton, respectively a and b, and X corresponds to undetected real radiation and beam remnants.\\
The cross section for single inclusive gluon production within HEF at leading order approximation  in $\alpha_s \ln1/x$  with two off shell initial state gluons (this formula is called also $k_T$ factorization) reads \cite{Gribov:1984tu,Kovchegov:2001sc}
\be
\label{cross section pT}
N(p_T,y)\equiv\frac{d\sigma}{d^{2}p_T dy} = \frac{2\alpha_s}{C_F}\frac{1}{p_T^2} \int d^{2} q_T {\cal F}_1^p(x_1,q_T^2,\mu^2) {\cal F}_2^p(x_2,(p_T-q_T)^2,\mu^2),
\ee
the formula applies in the situation where the system of gluons is dilute i.e. both of protons can be parametrized by gluon density coming from BFKL equation.
One can generalize the above formula to calculations when the nonlinearities are take into account in at least one of the colliding hadrons.
When the proton collides with lead the formula for single inclusive jet cross section in the central rapidity region reads
\be
\label{cross section pT}
N(p_T,y)\equiv\frac{d\sigma}{d^{2}p_T dy} = \frac{2\alpha_s}{C_F}\frac{1}{p_T^2} \int d^{2} q_T {\cal F}_1^p(x_1,q_T^2,\mu^2) {\cal F}_2^A(x_2,(p_T-q_T)^2,\mu^2),
\ee
here ${\cal F}_1^p(x_1,q_T^2,\mu^2)$ is the unintegrated gluon densities of the proton and  ${\cal F}_2^A(x_2,(p_T-q_T)^2,\mu^2)$ is the gluon density of nucleus in color adjoint representation. Both depend on longitudinal momentum fraction $x$, transversal momenta, and in general on factorization scale $\mu$.
The formula is valid when $x_1 > x_2$.
The longitudinal momentum fractions can be expressed in terms of rapidity-$y$ and transversal momentum-$p_T$ of jet produced in the final state as:
\be
x_1=\frac{p_T\,e^{y}}{\sqrt{S}},\,\,x_2=\frac{p_T\,e^{-y}}{\sqrt{S}}
\ee
where $\sqrt {S}$ is the total energy of the collision. In the situation when the jet is produced in the forward rapidity region the kinematics of the initial state is such that 
the $x$ values are very different. We consider the case when the produced forward jet goes along the proton direction; this corresponds to the situation such that $1\sim x_1>\!\!>x_2$. In such a regime one derives the so-called hybrid factorization formula valid which reads \cite{Dumitru:2005gt}
\begin{equation}
\begin{split}
\label{cross section hyb}
N(p_T,y)\equiv\frac{d\sigma}{dy_1d^2p_{1T}}= \frac{1}{2}\frac{1}{2(x_1x_2 S)^2}&\bigg[\sum_{q(\bar q)}\overline{|{\cal M}_{g^*q(\bar q)\to q(\bar q)}|}^2
   x_1f_{q(\bar q)/1}(x_1,\mu^2)\, {\cal F}_{g/2}^F(x_2,p_{1T}^2,\mu^2)\\
&+\overline{|{\cal M}_{g^*g\to g}|}^2x_1g_{g/1}(x_1,\mu^2)\, {\cal F}_{g/2}^A(x_2,p_{1T}^2,\mu^2))\bigg] \,.
\end{split}
\end{equation}
In both of the formulas (\ref{cross section pT}) and (\ref{cross section hyb}) the gluon density in the adjoint representation ${\cal F}^A(x,k^2)$ can be obtained from the fundamental one, using the gluon density in fundamental representation via the formulas collected in the Appendix. 
The parton densities $x_1f_{a/1}(x_1,\mu^2)$ are standard collinear parton densities and in our calculations we use CTEQ10NLO parton densities. 
The matrix elements $ \overline{|{\cal M}_{g^*g\to g}|^2}$, $\overline{|{\cal M}_{g^*q(\bar q)\to q(\bar q)}|}^2$ can be found in \cite{Bury:2016cue}.\\
The rapidity dependence of the mean transverse momentum of the jet can be calculated using the formula:
\be
\label{p_T}
\left\langle p_T \right\rangle = \frac{\int d^{2}p_T p_T  N(y,p_T)}{\int d^{2}p_T N(y,p_T)}, 
\ee
while the formulas after including the fragmentation function (FF) for $k_T$ factorization and hybrid factorization read:

\be
\nonumber
\label{p_T}
\left\langle p_T \right\rangle_{FF} = \frac{\int d^{2}p_T p_T \int_{0.05}^1 dz \frac{D_{g}(z)}{z^2} N(y,\frac{p_T}{z})}{\int d^{2}p_T\int_{0.05}^1 dz \frac{D_{g}(z)}{z^2}N_g(y,\frac{p_T}{z})}, 
\ee
\be
\left\langle p_T \right\rangle_{FF} = \frac{\int d^{2}p_T p_T \int_{\tau}^1 dz (\frac{D_{g}(z)}{z^2} N(y,\frac{p_T}{z}) + \sum_q\frac{D_{q}(z)}{z^2} N_q(y,\frac{p_T}{z})) }{\int d^{2}p_T\int_{\tau}^1 dz (\frac{D_{g}(z)}{z^2}N(y,\frac{p_T}{z})+\sum_q\frac{D_{q}(z)}{z^2}N(y,\frac{p_T}{z}))}, 
\ee
where $N(y, \frac{p_T}{z})$ is the double differential cross section for single inclusive  jet production and $D_{g}(z)$ is a fragmentation function describing probability for conversion of gluonic or quark jet to hadron \footnote{we consider pions which dominate the cross section} respectively and $\tau$ is allowed by the kinematics the lowest value of $z$. 
In this paper we use the fragmentation functions obtained in  \cite{Kniehl:2000fe}. In order to see the general trend and model dependence of the behavior of average $\left\langle p_T \right\rangle$ we calculate this observable using the following sets of unintegrated gluon distributions:

\begin{itemize}
	\item KS \cite{Kutak:2012rf} being a solution of momentum space version of BK equation with modifications according to the Kwieci\' nski-Martin-Satsto (KMS) prescription \cite{Kwiecinski:1997ee} which include kinematical constraint, complete splitting function and contribution of quarks and running coupling,
         \item KS-linear i.e. being a solution of momentum space version of the BFKL equation with modifications according to the KMS prescription described above,
	 \item KS-lead the same as KS but for lead nucleus. We use it to obtain predictions for $p$+$Pb$ initial configuration,
	
        \item gluon obtained using  the Kimber-Martin-Ryskin-Watt (KMRW) prescription. This is a gluon density obtained from collinear gluon density via resummation of virtual and soft emissions via the Sudakov exponent \cite{Kutak:2016mik}. The gluon density is valid in large and moderate values of $x$.
	\item Golec-Biernat-Wusthoff saturation model \cite{GolecBiernat:1998js} both for proton and for lead nucleus.
	
\end{itemize}

\section{Numerical results}
\label{sec-pPb-coll}

In the following sections we present the numerical results for $p$+$Pb$ and $p$+$p$ collision obtained in the $k_{T}$-factorization, hybrid factorization with fragmentation function and without fragmentation function. When we present the same factorization scheme for the different unintegrated gluon distribution function we use the different hue of the same color. In all other cases, we will use the following notation: 
\begin{itemize}

\item
red (solid) - $k_T$ factorization,

\item
blue (solid) - hybrid factorization without fragmentation function,

\item
purple (dashes) -  hybrid factorization with fragmentation function (FF).

\end{itemize}

As a first result of the paper we present comparison of the result for average $\left\langle p_{T}\right\rangle$ in $p$+$Pb$ collisions at $\sqrt{5}$ TeV as obtained in $k_T$ factorization with and without accounting for fragmentation function. In the $k_T$-factorization calculation we are going to use for the lead nuclei the KS-lead unintegrated parton density function, while for the hadron with partons having larger $x$ other unintegrated gluon densities from the list presented in the previous section. The crucial point is the appropriate choice of cuoff i.e. lower limit for integration over the transversal momenta. We choose it 1.3 GeV, as dictated by the lowest momentum scale of the collinear parton density used in hybrid factorization. Choice of lower cutoff which is possible in $k_T$ factorization would make the results hard to be compared.   
In the hybrid factorization calculations the parton density of lead is going to be parametrized by the KS-lead unintegrated gluon density or GBW \footnote{with modified saturation scale to account for higher density of gluons} gluon density while the parton density of large $x$ proton is provided by CTEQ10NLO collinear parton density functions.

\begin{figure}[t]
\hspace{-9mm}
\begin{minipage}{0.5 \textwidth}
\center
\includegraphics[width=\textwidth]{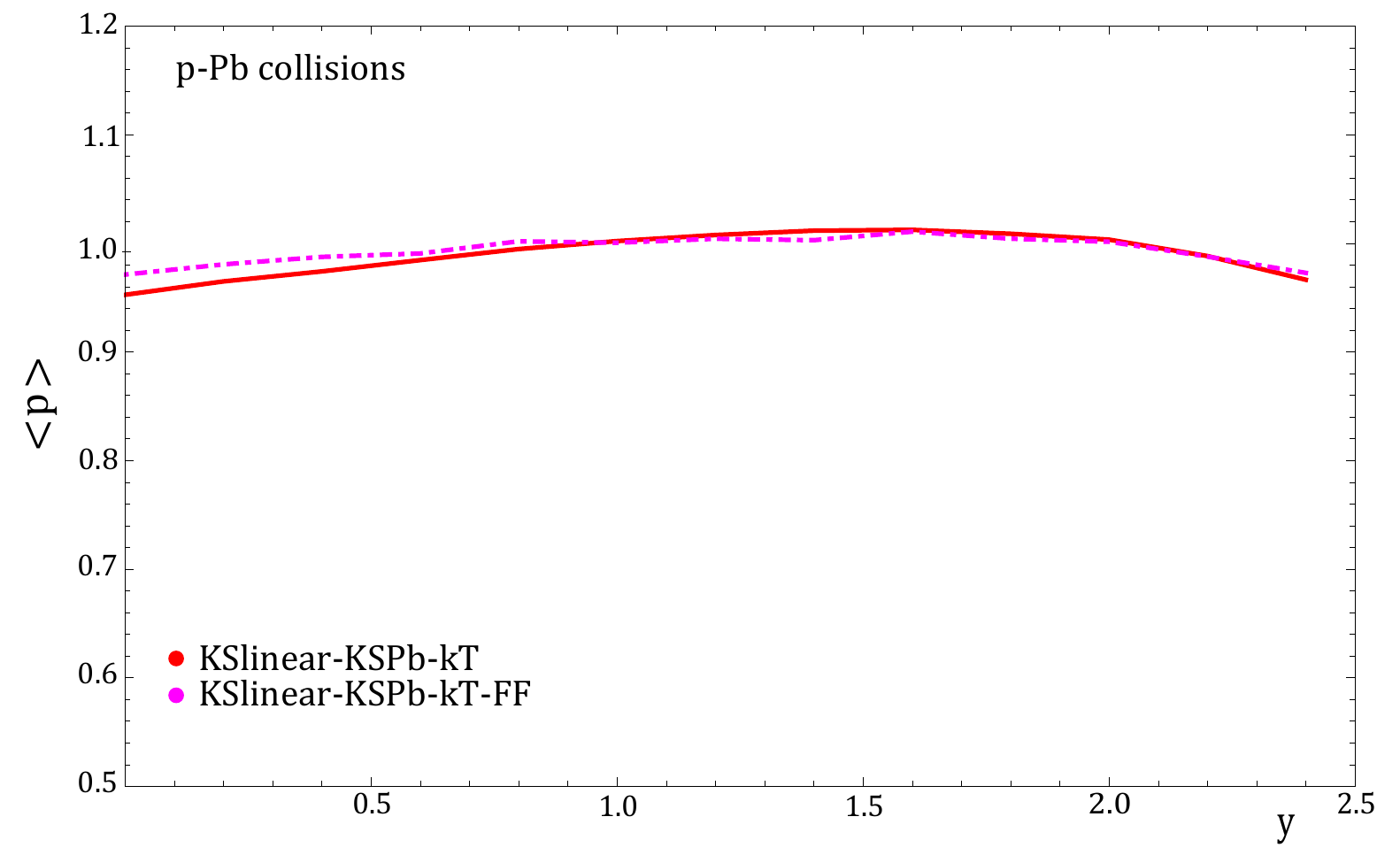}
\vspace{-2mm}
\caption{\small{(Colour online) The rapidity dependence of the ratio $\left\langle p_{T}(y,\sqrt{s})\right\rangle/\left\langle p_{T}(2,\sqrt{s})\right\rangle$ in p-Pb collisions as obtained using $k_{T}$ factorization. The Parton Distribution Function PDF of proton is given by KS-linear unintegrated gluon density while the PDF of lead in colour adjoint representations is given by KS-nonlinear unintegrated gluon density.}}
\label{fig1}
\end{minipage}
\hspace{2mm}
\begin{minipage}{0.5 \textwidth}
\center
\includegraphics[width=\textwidth]{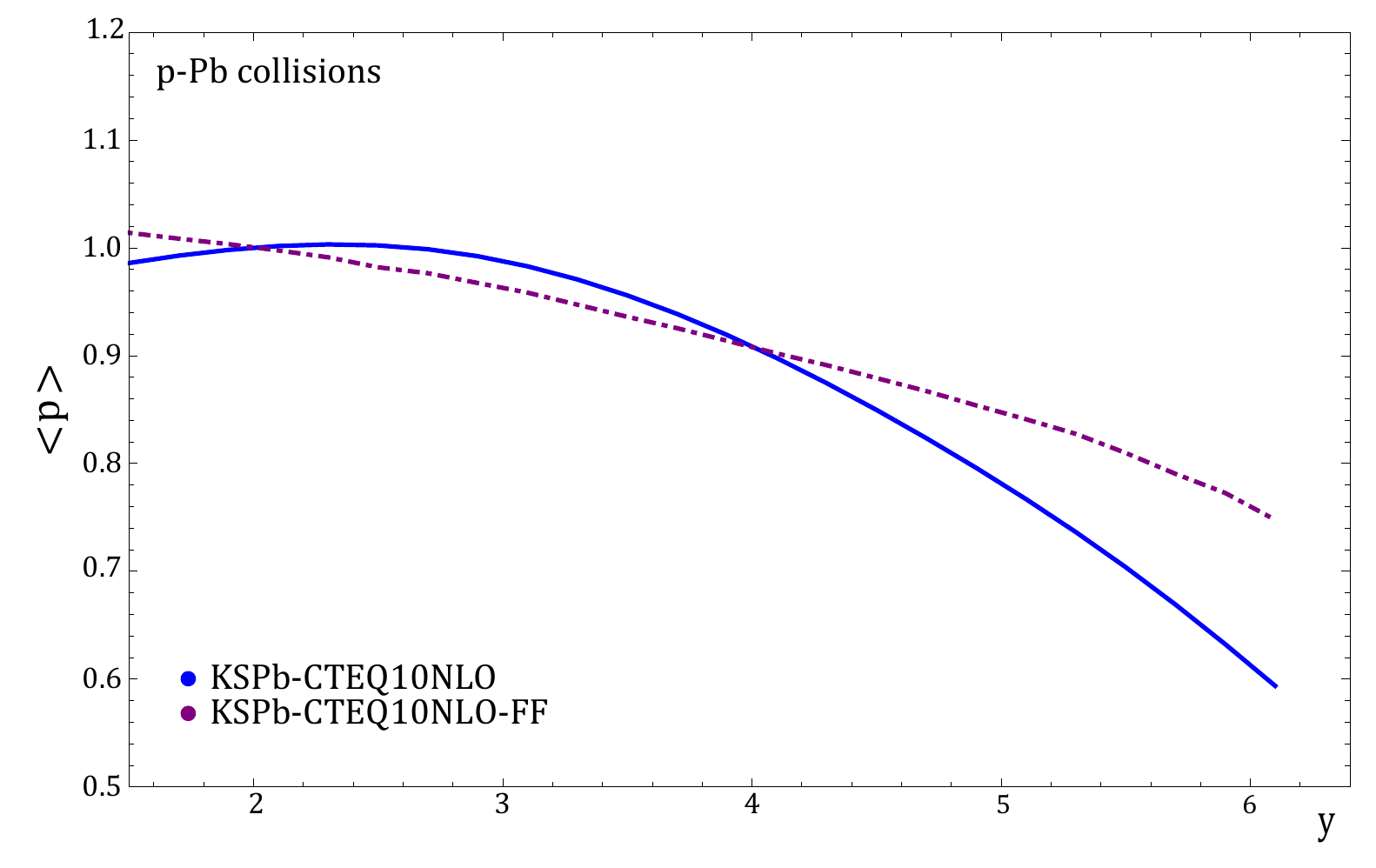}
\vspace{1mm}
\caption{\small{(Colour online) The rapidity dependence of the ratio $\left\langle p_{T}(y,\sqrt{s})\right\rangle/\left\langle p_{T}(2,\sqrt{s})\right\rangle$ in p-Pb collisions obtained in hybrid factorization. 
The PDF for lead is given by KS-nonlinear unintegrated gluon density, while for proton we take CTEQ10NLO set of PDFs.}}
\label{fig2}
\end{minipage}
\end{figure}

In Fig.(\ref{fig1}) we see that the result of the evaluation of the $\left\langle p_{T}\right\rangle$ (normalized to value at rapidity $y=2$) within the $k_T$ factorization approach weakly depends on whether the fragmentation function (FF) is included or not, this result was argued to hold already by \cite{Bozek:2013sda}. To arrive at the result the authors of the mentioned paper used general arguments and simplified analytical calculations within GBW-like saturation model. The small difference between the two curves obtained can be shown to come from the running coupling of $\alpha_S$. In Fig.(\ref{fig2}) we present the result obtained within hybrid factorization, here the difference between particle spectrum and jet spectrum is clearly visible. The cross section for evaluation of jets is slowly increasing and then decreases at forward rapidities while the particle spectrum decreases monotonously. This results are therefore consistent with the one obtained in \cite{Duraes:2015qoa}. At this point we would like to stress that for this calculation we used the unintegrated gluon density of proton coming from the linear evolution equation therefore all saturation effects visible for jets are coming from $Pb$ nucleus. \\

\begin{figure}[t]
\hspace{2mm}
\begin{minipage}{0.5 \textwidth}
\centering
\includegraphics[width=1.0\textwidth]{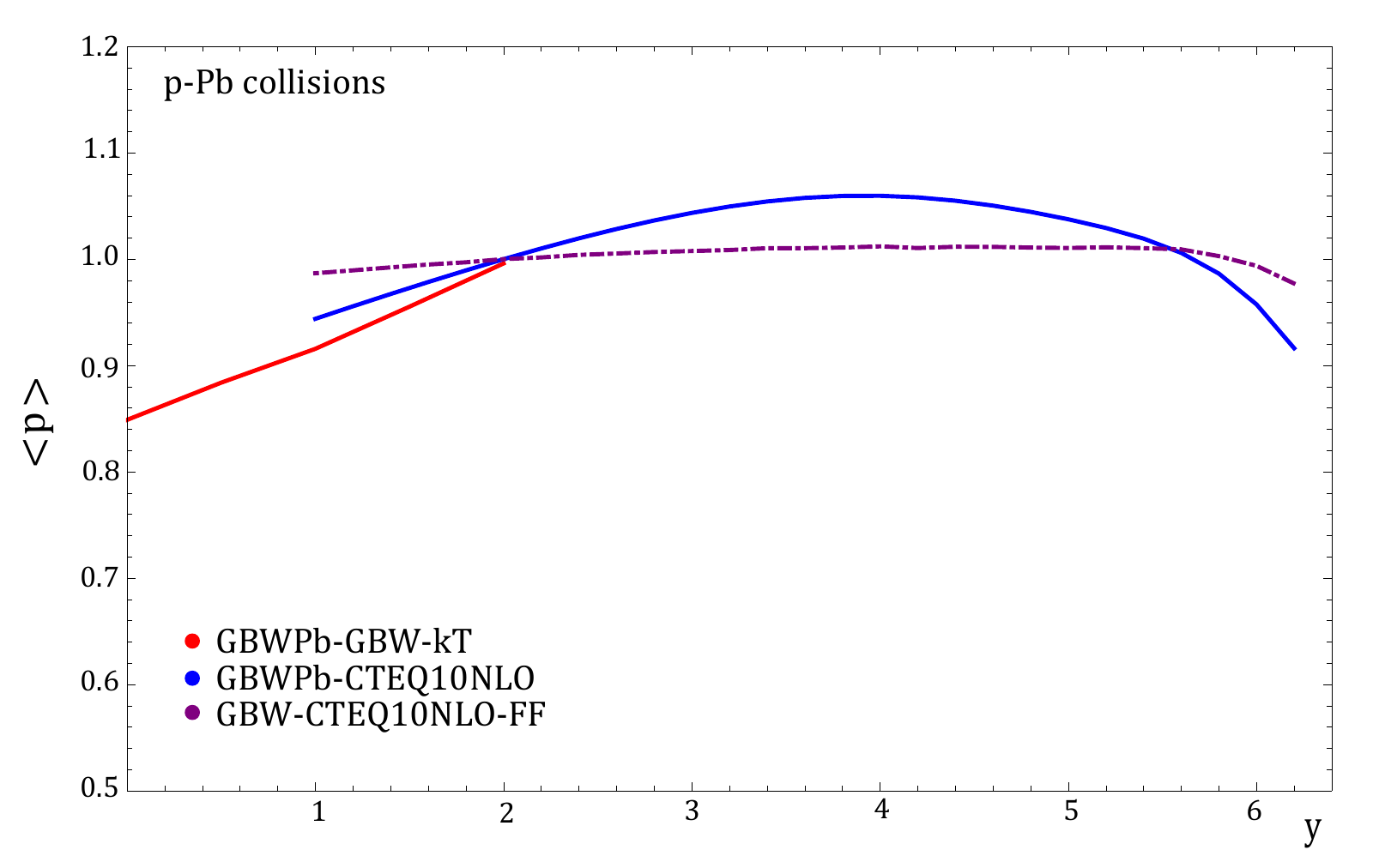}
\vspace{-2mm}
\caption{\small{(Colour online) The rapidity dependence of the ratio $\left\langle p_{T}(y,\sqrt{s})\right\rangle/\left\langle p_{T}(2,\sqrt{s})\right\rangle$ in $p$-Pb collisions as obtained using $k_{T}$ or hybrid factorization. The PDF of proton and lead is obtained using the GBW model.}}
\label{fig3}
\end{minipage}
\hspace{2mm}
\begin{minipage}{0.5 \textwidth}
\centering
\includegraphics[width=1.0\textwidth]{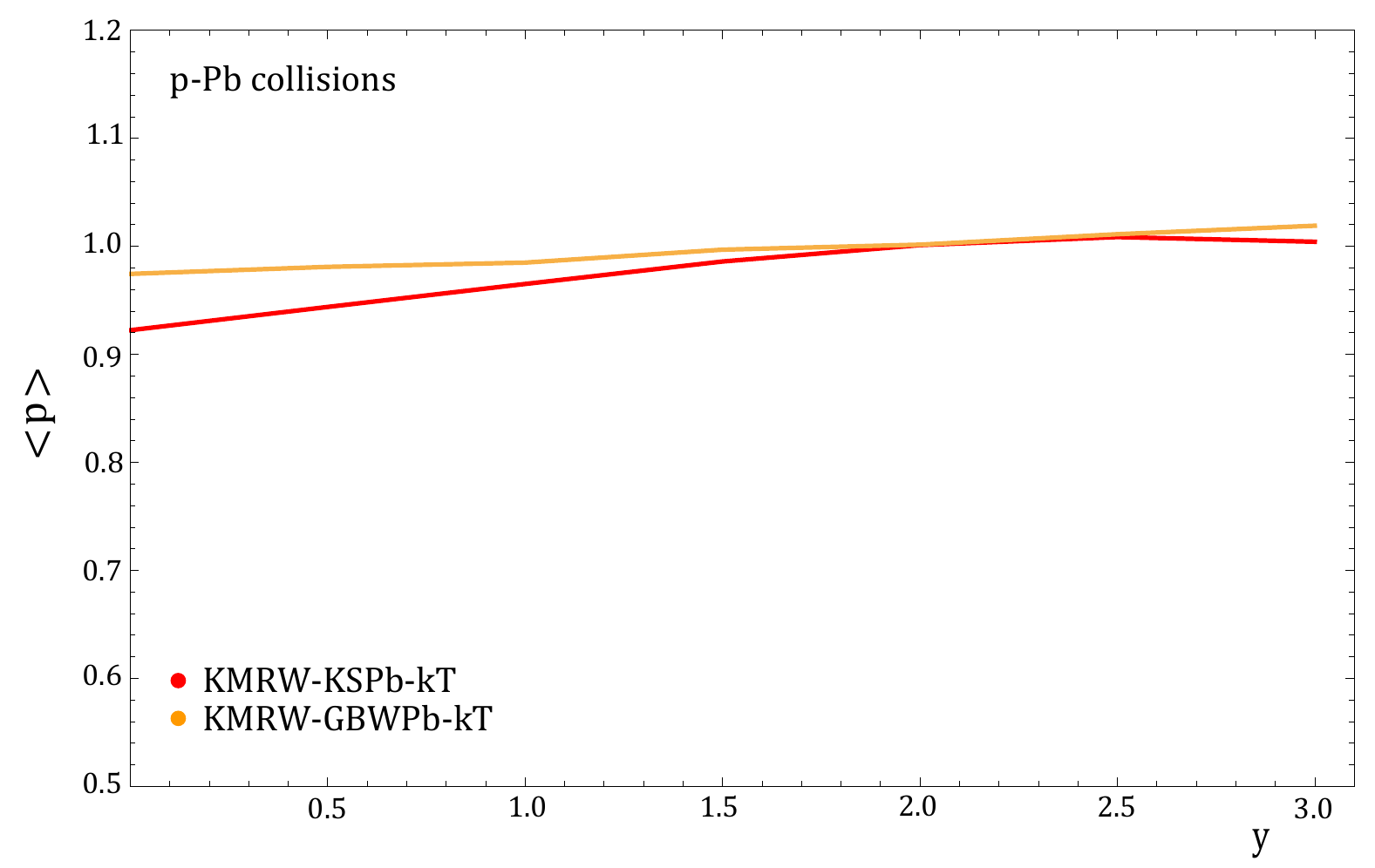}
\vspace{-3mm}
\caption{\small{(Colour online) The rapidity dependence of the ratio $\left\langle p_{T}(y,\sqrt{s})\right\rangle/\left\langle p_{T}(2,\sqrt{s})\right\rangle$ in p-Pb collisions obtained in $k_{T}$-factorization. Unintegrated gluon density for proton was obtained using KMRW prescription and for lead GBW saturation (orange curve) and KS-nonlinear (red curve) was used.}}
\label{fig4}
\end{minipage}
\end{figure}

Fig.(\ref{fig3}) shows the comparison of the results obtained in $k_T$ factorization with those obtained in hybrid factorization using GBW and the GBW-lead model to parametrize the proton and the nucleon respectively. In hybrid factorization only the lead nucleus is parametrized by the GBW model. The main difference in the result as to the one obtained using KS-gluon density is that the GBW decays exponentially towards large $k_T$ and therefore the observable is more dominated by saturation scale. We skip the result of $k_T$ factorization that includes FF since it is essentially the same as without it.
Another interesting point to study is to investigate the difference in the $\langle p_T\rangle$ spectrum when the $Pb$ nucleus is parametrized by GBW or BK parton densities, while the large $x$ proton by the KMRW unintegrated gluon density which is derived from collinear physics with appropriate resummation. We see in Fig.(\ref{fig4}) that in the region where the framework is applicable the results are very similar and suggest that the low $k_T$ region dominates where the parton densities are essentially similar i.e. behave like $\sim k_T^2/Q_s^2$.

\begin{figure}[t!]
\hspace{-7mm}
\begin{minipage}{0.5 \textwidth}
\center
\includegraphics[width=1.0\textwidth]{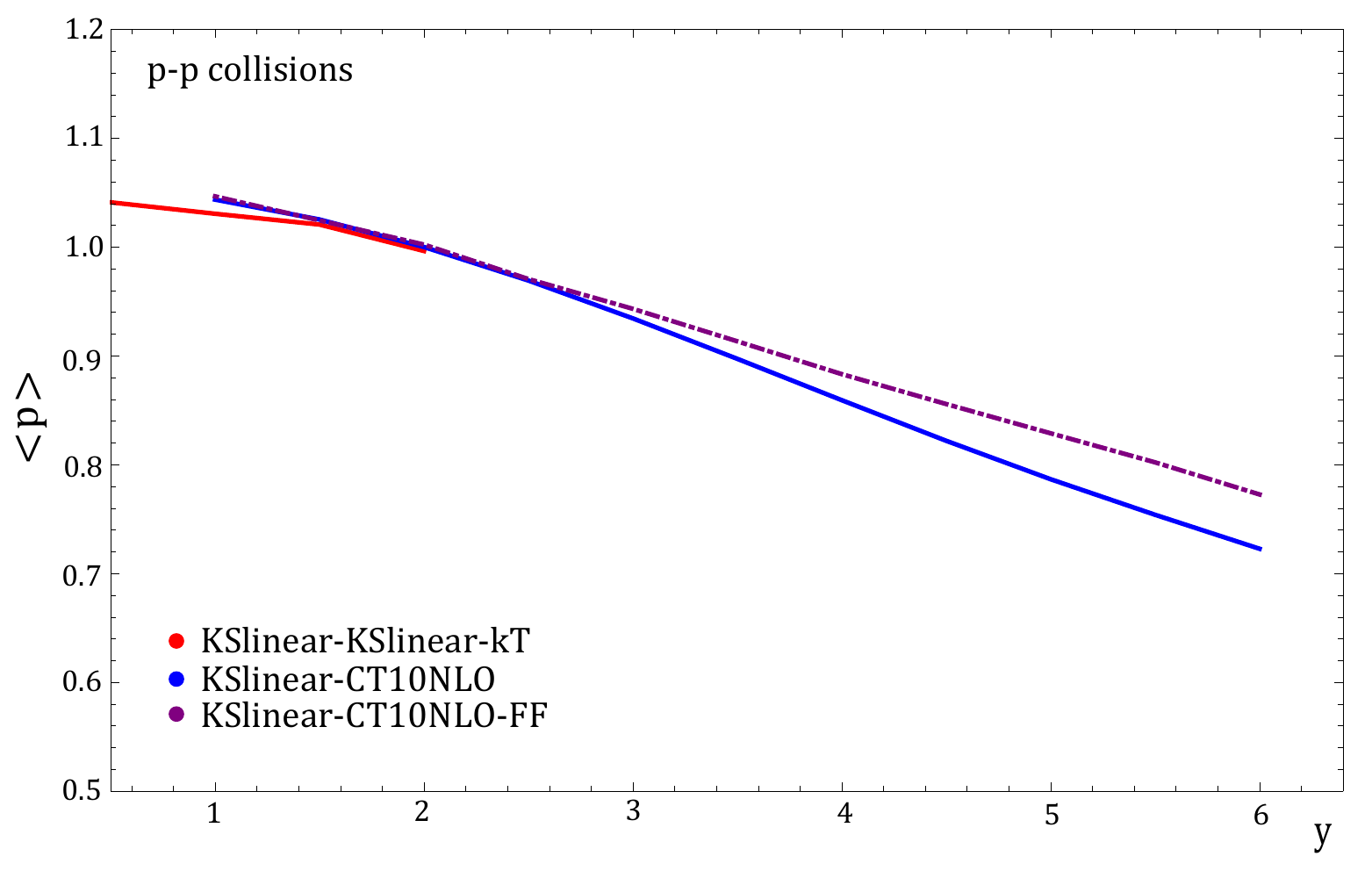}
\vspace{-2mm}
\caption{\small{(Colour online) The rapidity dependence of the ratio $\left\langle p_{T}(y,\sqrt{s})\right\rangle/\left\langle p_{T}(2,\sqrt{s})\right\rangle$ in $p$-$p$ collision. The gluon density parametrising proton was chosen to be KS-linear.}}
\label{fig5}
\end{minipage}
\hspace{2mm}
\begin{minipage}{0.5 \textwidth}
\center
\includegraphics[width=1.0\textwidth]{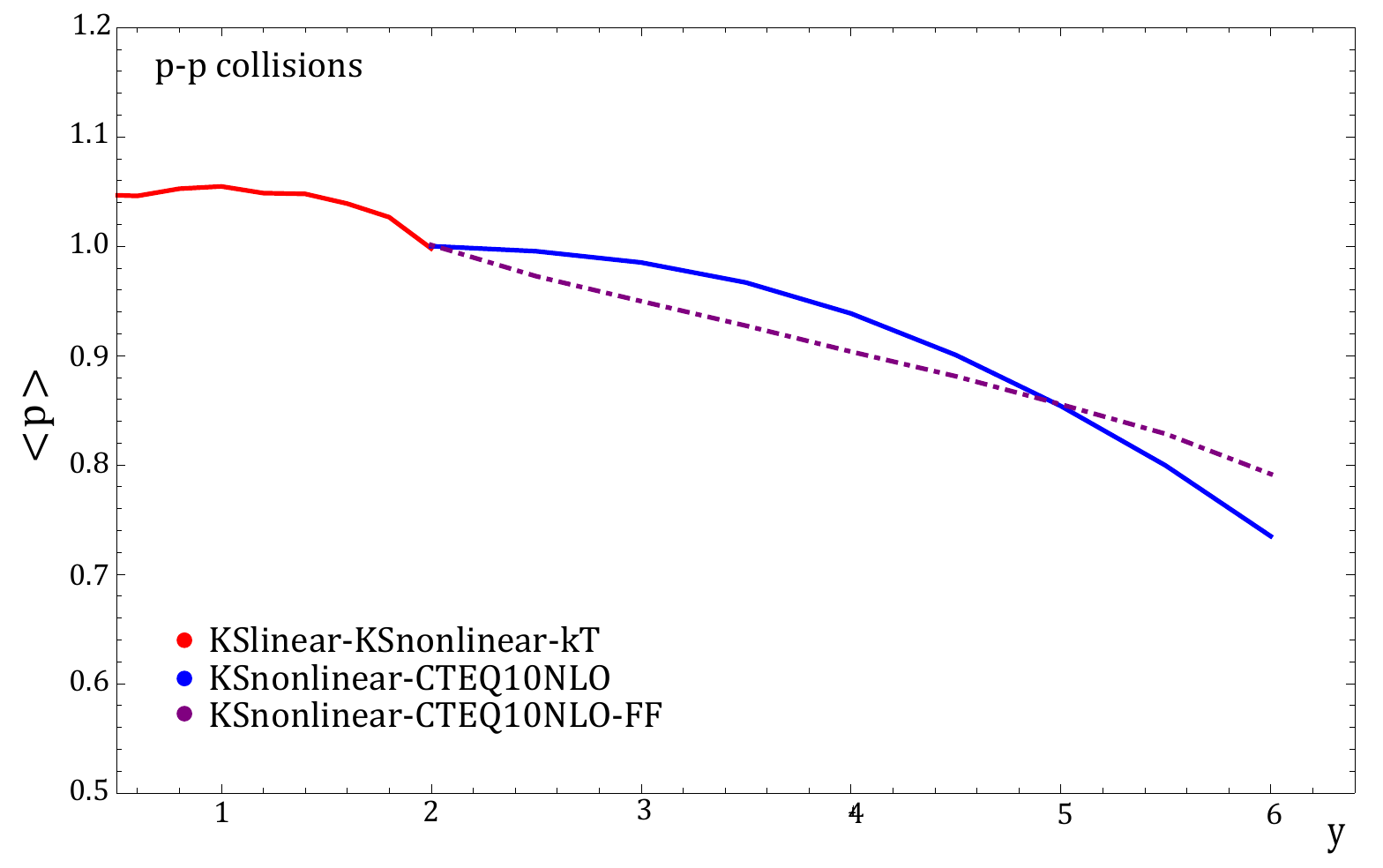}
\vspace{-3mm}
\caption{\small{(Colour online) The rapidity dependence of the ratio $\left\langle p_{T}(y,\sqrt{s})\right\rangle/\left\langle p_{T}(2,\sqrt{s})\right\rangle$ in $p$-$p$ collision. The gluon density parametrising proton with lower-$x$ was chosen to be KS-nonlinear.} }
\label{fig6}
\end{minipage}
\end{figure}

\begin{figure}[t!]
\hspace{-7mm}
\begin{minipage}{0.5 \textwidth}
\center
\includegraphics[width=1.0\textwidth]{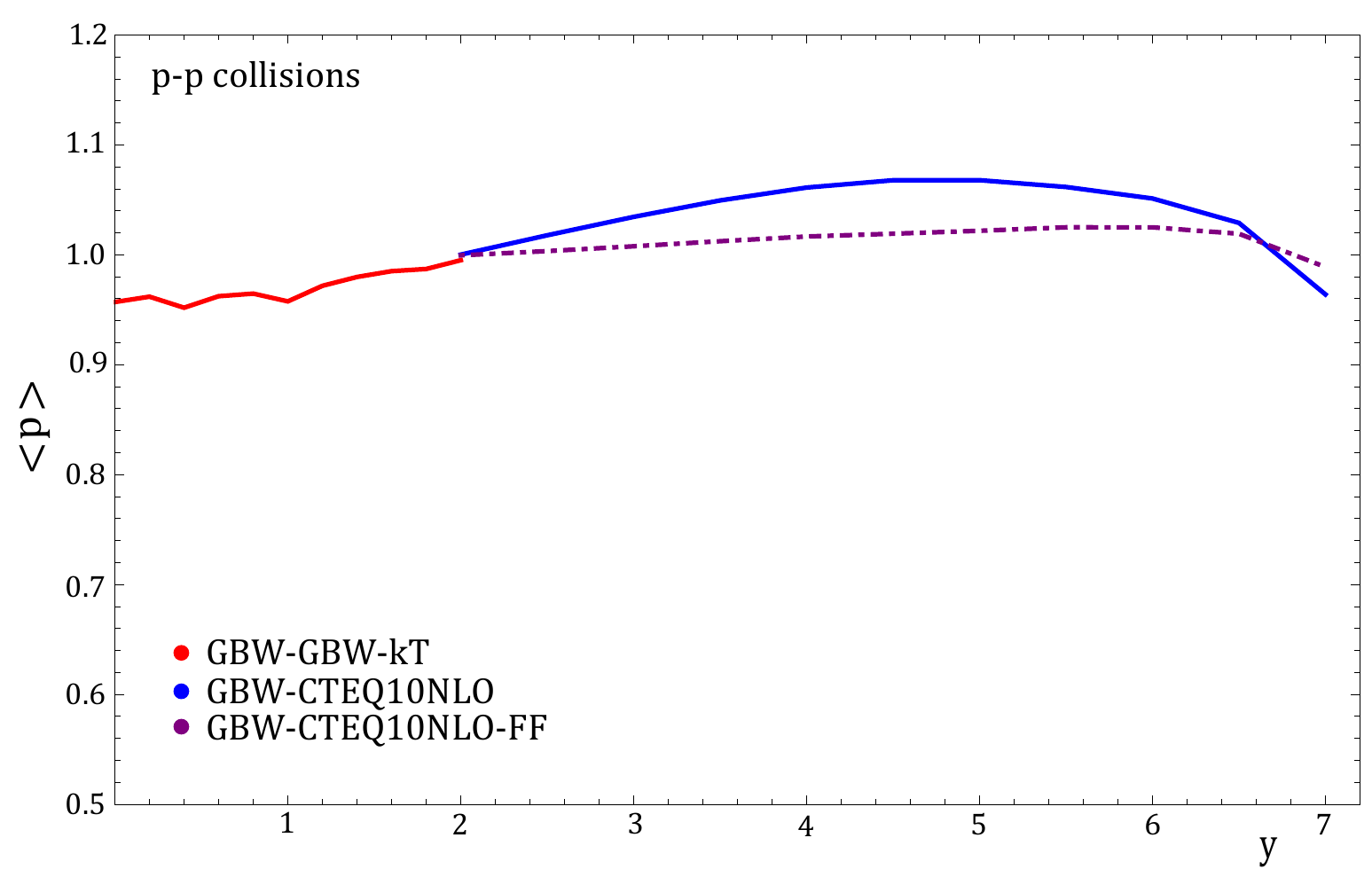}
\vspace{-2mm}
\caption{\small{(Colour online) The rapidity dependence of the ratio $\left\langle p_{T}(y,\sqrt{s})\right\rangle/\left\langle p_{T}(2,\sqrt{s})\right\rangle$ in p-p collision. The gluon density parametrizing proton was chosen to be parametrized by the GBW model.}}
\label{fig7}
\end{minipage}
\hspace{2mm}
\begin{minipage}{0.5 \textwidth}
\center
\includegraphics[width=1.0\textwidth]{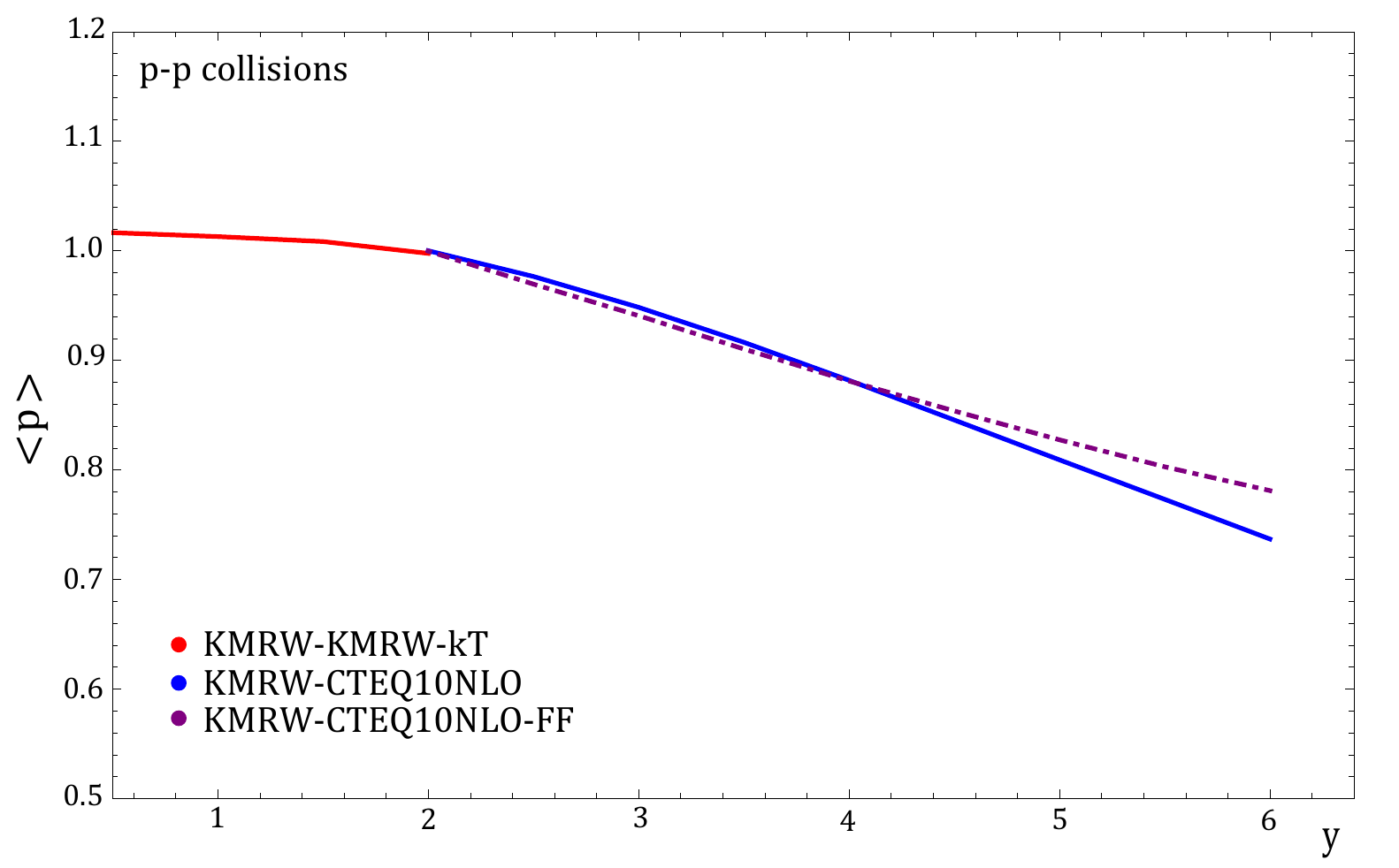}
\vspace{-1mm}
\caption{\small{(Colour online) The rapidity dependence of the ratio $\left\langle p_{T}(y,\sqrt{s})\right\rangle/\left\langle p_{T}(2,\sqrt{s})\right\rangle$ in p-p collision. The gluon density used is obtained within KMRW framework.}}
\label{fig8}
\end{minipage}
\end{figure}

We move towards the $p$-$p$ case; in Fig.(\ref{fig5})-Fig.(\ref{fig8}) we investigate the dependence of $\langle p\rangle$ on rapidity in $p$+$p$ collisions. The general conclusion is that for $p$+$p$ collisions there is not much difference between results with saturation and without saturation. The difference between results with and without the fragmentation function is however bigger in cases where parton density with saturation for at least one proton was used. Furthermore comparing purple curves obtained in Fig.(\ref{fig5}) and Fig.(\ref{fig6}) one sees that they do not differ much so the particle spectra are not optimal to distinguish between the saturation and the no saturation case.
One can however see a significant difference to the result obtained within the GBW saturation model which predicts in the large region of rapidity an increase of the $\langle p\rangle$. In the $p$+$Pb$ case the results where the saturation scale in $Pb$ is large it leads to an increase of mean transversal momentum in central and forward values of rapidities. 
The general trend in behavior of frameworks with and without saturation in $p$-$p$ collisions can be visualized by comparing figures obtained using KS-linear (Fig.\ref{fig5}), KMRW (Fig.\ref{fig8}) with those where KS-nonlinear (Fig.\ref{fig6}) or GBW (Fig.\ref{fig7}) was used, it is visible that in the later case $\langle p \rangle$ increases as a function of rapidity (although in the KS-linear case the distribution is almost flat in the central rapidity region). 

To conclude with the results obtained above  we see that the behavior of the average transversal momentum as obtained above behaves as follows:
\begin{itemize}
\item it grows in the central rapidity region if one assumes saturation (although very weakly in the $p$+$p$ case), 
\item it grows for $2<y<3$ and falls for $y>3$ for jets in $p$+$Pb$ collision if one assumes saturation,
\item is constant for jets in the region $2<y<3$ and falls for $y>3$ in the $p$+$p$ collision if one assumes saturation,
\item it falls for jets in the $p$+$p$ collision if one does not assume saturation,
\item it falls in all rapidity region for particles in $p$+$Pb$ and $p$+$p$ independent if one assumes saturation or not,
\item we used also the GBW model which has strong saturation effects. In this case the conclusion is that
the GBW model predicts a strong increase with rapidity of the average transversal momentum.
\end{itemize}

\begin{figure}[t]
\hspace{-7mm}
\begin{minipage}{0.5 \textwidth}
\center
\includegraphics[width=1.0\textwidth]{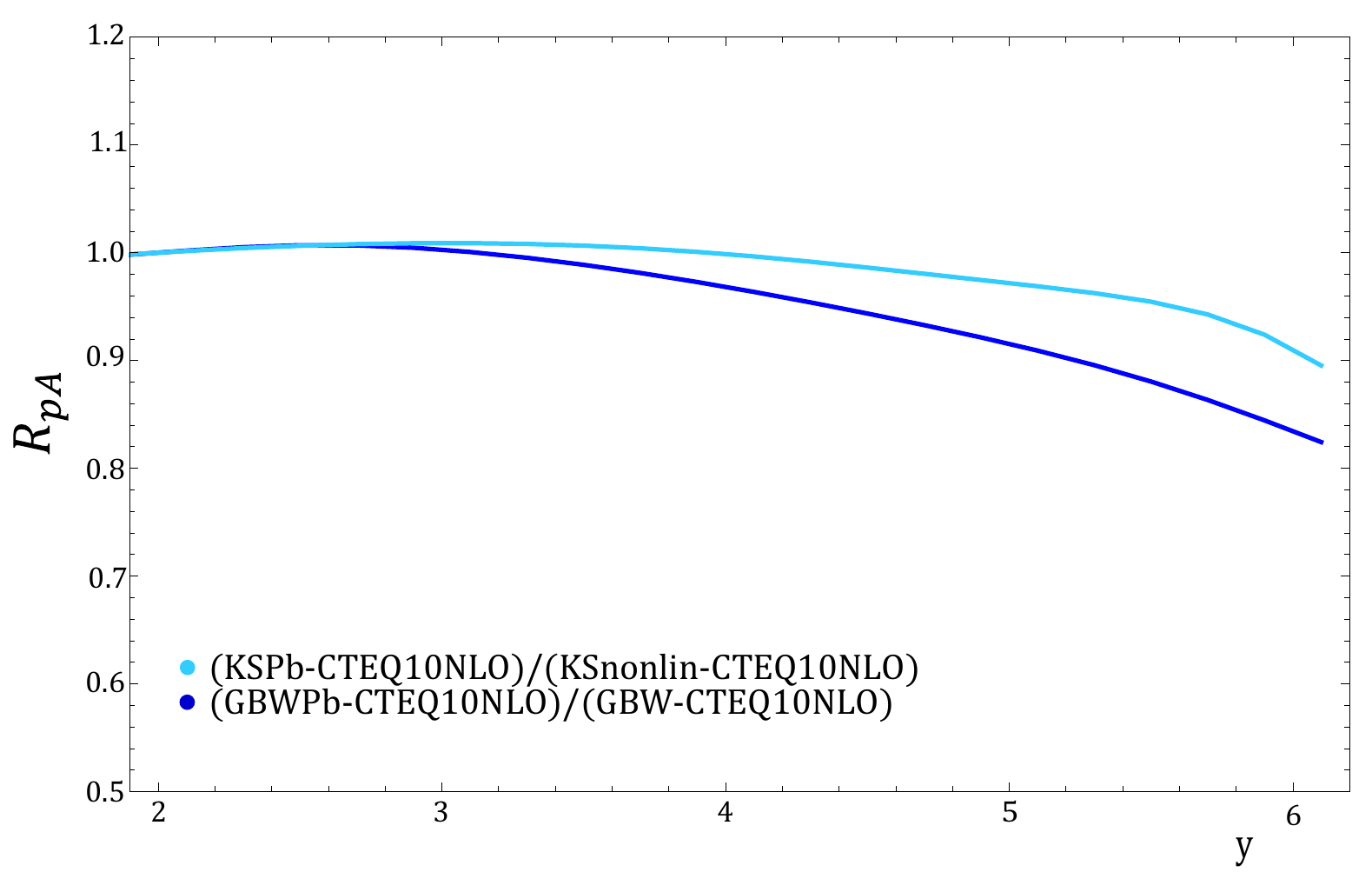}
\vspace{-2mm}
\caption{The nuclear modification ratio obtained using KS-nonlinear gluon density or GBW model for proton and lead. }
\label{fig9}
\end{minipage}
\hspace{2mm}
\begin{minipage}{0.5 \textwidth}
\center
\includegraphics[width=1.0\textwidth]{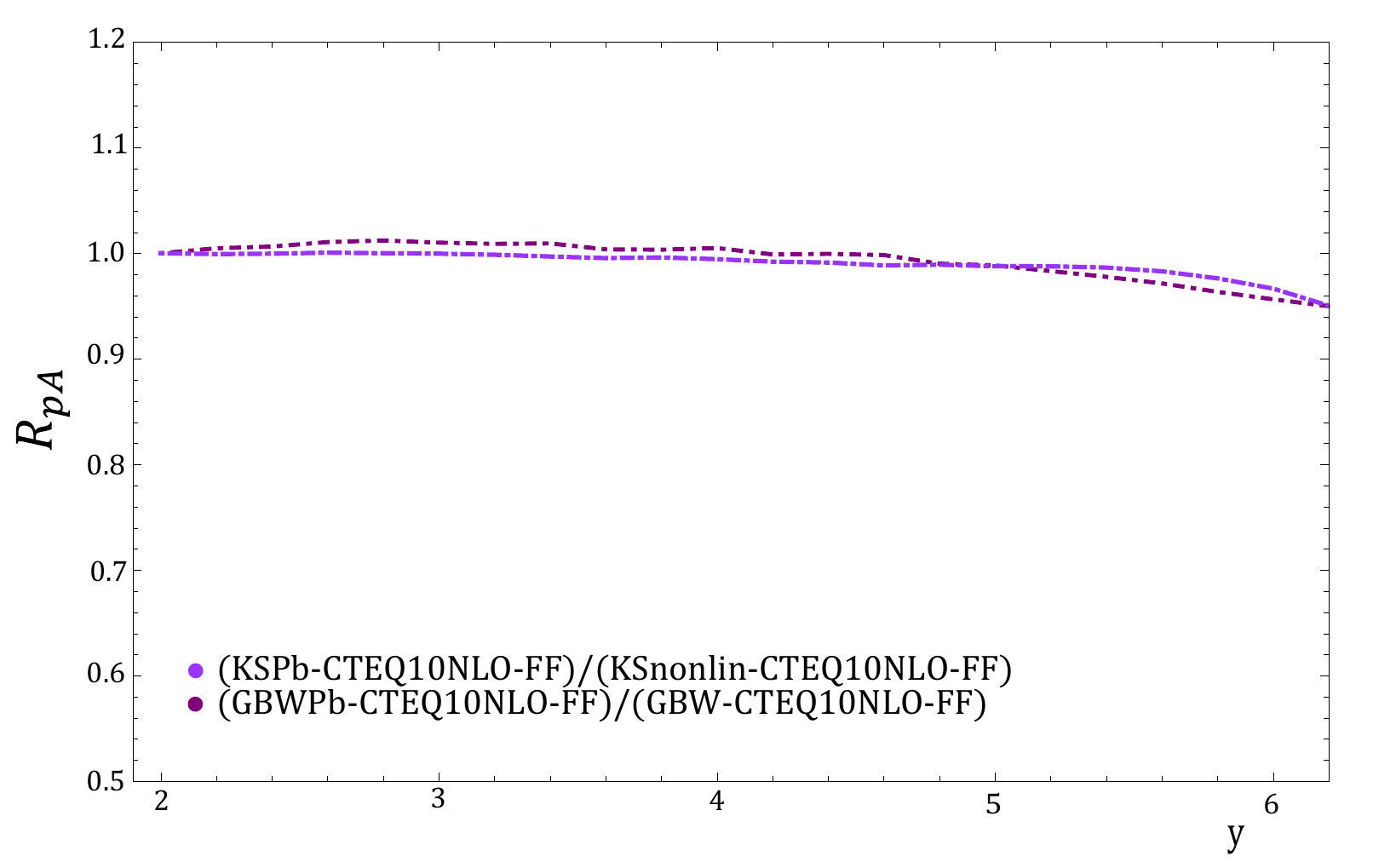}
\vspace{-1mm}
\caption{\small{The nuclear modification ratio obtained using KS-nonlinear gluon density or GBW model for proton and lead.}}
\label{fig10}
\end{minipage}
\end{figure}
In order to measure the effects of increased density of partons as one goes from $p$+$p$ to $p$+$Pb$ scattering one often measures so-called nuclear modification ratio $R_{pA}$.
This quantity is defined as
\begin{equation}
R_{pA}=\frac{\frac{d\sigma^{p+A}}{dO}}{\frac{A d\sigma^{p+p}}{dO}}
\end{equation}
with $A=208$ for $Pb$. In the region of phase space where the saturation effects are weak, this ratio is equal to unity.
However, in the more forward rapidity region the nonlinear evolution plays a more important role in the case of the nucleus $R_{pPb}$ and therefore the ratio goes below 1. We see in Fig. (\ref{fig9}) that at the forward region there is suppression of the cross section for jets. However, as one goes from jets to particles the result is consistent with unity, see Fig.(\ref{fig10}).

\section{Conclusions}
\label{sec-conclusions}
In the paper we studied average transversal momentum of single inclusive jet and particle produced in $p$+$p$ and $p$+$Pb$ collisions. 
The study has been performed within the framework of high energy factorization. 
We see that as soon as we use at least for one initial state hadron parton density with saturation there is region in rapidity where
the transversal momentum of produced jet is growing.. However, when the parton densities do not include saturation effects the distribution is essentially falling both for jets and particles in all rapidity regions. 
We therefore confirm the result by \cite{Bozek:2013sda}, that once the parton density function with saturation is used the average transversal momentum grows.
However, the application of the $k_T$ factorization approach used by \cite{Bozek:2013sda} should be limited to central rapidity region $y\le 2$ where the longitudinal momenta of initial state partons are comparable. As one goes to the forward rapidity region the hybrid high energy factorization is the proper framework to be used as has been done in \cite{Duraes:2015qoa}.
Besides revisiting the results of the mentioned groups we performed calculations using parton densities which do not include saturation effects. We observed that in such case the 
average transversal momentum is always decreasing with increasing rapidity independent on the factorization used nor final state, i.e. jet or particle.
This allows one to conclude that indeed saturation predicts an increase of transversal momentum of jet or particle as a function of rapidity in a certain rapidity window.
Finally we studied the nuclear modification ratio of forward jets and particles, and we see that the suppression of the average transversal momenta  is stronger for jets than for particles.

\section*{Acknowledgments}
We would like to thank Krzysztof Golec-Biernat for useful comments. The work of K.K. has been supported by Narodowe Centrum Nauki
with Sonata Bis Grant No. DEC-2013/10/E/ST2/00656.

\section{Appendix: Gluon density in color adjoint representation}
The dipole amplitude in adjoint representation expressed in terms of fundamental representation reads \cite{JalilianMarian:2005jf}:
\begin{equation}
{\cal N}^A(x,r)=2{\cal N}^F(x,r)-({\cal N}^F(x,r))^2
\end{equation}
and it is related to the dipole amplitude in the color fundamental representation via
\begin{equation}
{\cal N}^F(x,r)=8 \frac{4\pi^3}{N_c\alpha_s S_\perp}\int\frac{dk}{k}[1-J_0(kr)]{\cal F}^F(x,k^2)
\end{equation}
where  ${\cal N}^F(x,r)$ can be provided by the solution of the BK or JIMWLK evolution equation or model. In the above $r\equiv {\bf r}$ is a two-dimensional vector denoting relative distance of end points of the dipole whose modulus represents size of a color dipole.
The gluon density in the color adjoint representation is calculated via:
\begin{equation}
{\cal F}^A(x,k^2)=S_{\perp}\frac{N_c}{4\alpha_s\pi^2}k^2\int\frac{d^2r}{2\pi}e^{-i\,k\cdot r}(1-{\cal N}^A(x,r))
\end{equation}

%

\end{document}